\documentstyle[preprint,aps]{revtex}

\begin{document}

\title{Primordial Nucleosynthesis as a test of variable rest masses 
5-dimensional cosmology}

\author{Luis A. Anchordoqui\thanks{E-mail: doqui@venus.fisica.unlp.edu.ar}, 
Diego F. Torres, H\'ector Vucetich}
\address{Departamento de F\'{\i}sica - Universidad Nacional de La Plata\\
C.C. 67 - C.P. 1900 - La Plata -$ Argentina$}

\maketitle

\begin{abstract}
The deviation of primordial Helium production due 
to a variation on the difference between the rest masses of the nucleons is 
presented. It is found an upper bound $\delta (M_{_n} - M_{_p}) 
\alt 0.129$ MeV, between the present and nucleosynthesis epochs. 
This bound 
is used to analyze Wesson's theory of gravitation; as a result, it is ruled 
out by observation.

\vspace{1cm}


\end{abstract}
\newpage
The so-called Hot Big Bang model provides a consistent
description of the evolution of the Universe. This model depends 
on a set of universal parameters known as {\em fundamental constants}.
However, a great number of theories of gravitation have appeared 
in which some of these {\em
constants} do vary with time. In particular, scalar-tensor theories
of gravity \cite{BD,Will,Barrow} make predictions which are in complete
agreement with present experimental data while predicting a variation 
of Newton's 
gravitational constant over cosmological time scales. The physical 
embodiment of these theories allows a natural generalization of General 
Relativity and thus provides a convenient set of representations for the
observational limits on possible deviations of Einstein's theory,
making them a profitable arena for cosmology.

The hypothesis that
the gravitational interaction has changed over the history of the
Universe (i.e. the gravitational parameter $G$ or the rest masses of 
elementary particles depend on the time $t$) can be also analyzed 
in the framework of a 5-dimensional cosmology, proposed by Wesson 
\cite{WES}. This theory is founded either on dimensional analysis 
\cite{{WM,Ma}} as well as on
reinterpretation of the five dimensional vacuum equations\cite{Wes,Wess}.
A consequence of this relation is that the 
rest 
mass of a given
body varies from point to point in space-time, in agreement with the 
ideas of
Mach\cite{Mach,Soleng}.
This is a definite and testable prediction, especially when time
intervals of cosmological order are considered \cite{WES,WESS}.

In this letter, we find an upper bound to the 
variation of the masses of the nucleons over cosmological time intervals, 
from a comparison of the observed
primordial abundance of $^4$He with the theoretical variation induced by a
changing mass. Subsequently, this upper bound is compared with the
prediction of the 5--dimensional Wesson's theory of gravitation.

\vspace{.7cm}

If we consider that a variation of the rest masses of the particles had
occurred between the epoch of primordial nucleosynthesis and ours, we can
compute the deviation in the $^4$He production from the Hot Big Bang
model prediction due to this fact. The method
we are going to apply  is a generalization of the 
calculation
made by Casas, Bellido and Quir\'os \cite{Casas} to fix nucleosynthesis
bounds on the variation of the gravitational constant in Jordan-Brans-Dicke
theory of gravity. The same method was recently used to limit other
scalar-tensor theories with more general couplings functions
\cite{Torres}. We know that -- in the Hot Big Bang model -- the
primordial $^4$He production is given by

\begin{equation}
\label{1}Y_{\rm p}=\lambda \left( \frac{2x}{1+x}\right) _{t_{\rm f}} 
\end{equation}
where \mbox{$\lambda = 
\exp (-(t_{\rm nuc}-t_{\rm f})/\tau )$} stands for the fraction of
neutrons which decayed into protons between $t_{\rm f}$ and $t_{\rm
nuc}$, with $t_{\rm f}$ $(t_{\rm nuc})$ the time of freeze out of the weak 
interactions
(nucleosynthesis) \cite{Kolb}, 
$\tau$
the neutron mean lifetime, and $x=\exp (-(M_n-M_p)/kT)$ the neutron to proton
ratio. If we make a variation on the rest masses of the particles, i.e. if
we consider that a difference between the rest masses in the
present and nucleosythesis epochs does exist, 
it is straightforward to compute an expression
for the deviation in the $^4$He primordial production, that reads
\begin{equation}
\label{2}\delta Y_{\rm p}=Y_{\rm p} \ln \left( \frac{2\lambda
}{Y_{\rm p}} - 1 \right) \left[ -1 + 
\frac{Y_{\rm p}}{2\lambda }\right] \frac{\delta (\Delta Q)}{\Delta Q} 
\end{equation}
where we have defined $\Delta Q=M_n-M_p$. 

We must also take into account that
a variation in the rest masses of the particles will affect the
neutron lifetime. The latter fact was not considered above since (\ref{2})
represents only the explicit derivative with respect to $\Delta Q$.
A calculation of how a variation of the neutron lifetime affects
the prediction on primordial $^4$He has been already done in \cite{Casas,Yp}. 
It is given by,
\begin{equation}
\label{2'}\delta Y_{\rm p}=0.185 \frac{\delta\tau}{\tau}
\end{equation}
Noting that the dependence of $\tau$ upon the masses is $\tau \propto
G^{-2}_{\rm F} \, \Delta Q^{-5} \propto M^4_{W} \, \Delta Q^{-5}$, where
$G_{\rm F}$ is the Fermi constant and $M_{W}$ is the mass 
of the bosonic mediator of weak interactions, it is easy to obtain,
\begin{equation}
\label{2''}\frac{\delta\tau}{\tau}=-\frac{\delta\Delta Q}{\Delta Q}
\end{equation}

Thus, any variable rest mass
theory will predict a primordial Helium abundance given by,
\begin{equation}
Y_{\rm p,var-rest-mass} = Y_{\rm p,std} + \delta Y_{\rm p}
\end{equation}
where $Y_{\rm p,std}$ is the value predicted by the standard big--bang
nucleosynthesis theory. 

We can find an upper bound for $\delta Y_{\rm
p}$ summing up the two deviations
referred above, {\it i.e.} equations (\ref{2}) and (\ref{2'}),
and comparing with the observed value $Y_{\rm obs}$,
\begin{equation}
| \delta Y_{\rm p} | \leq | Y_{\rm obs} - Y_{\rm std} | + \epsilon
\leq \sigma
\end{equation}
where $\epsilon$ is an estimate of the observational error and
$\sigma$ includes also estimates of theoretical errors. From \cite{Yp,pdg}
we estimate $\sigma \leq 0.01$. But, it may also be possible that, due 
to small 
changes in nucleon masses, small
changes in nuclear cross sections do occur. Since the functional dependence
of cross sections with masses is generally unknown, we shall take into account
these changes by arbitrarily doubling the theoretical error.
Thus, we obtain
\begin{equation}
\label{4}\left|\frac{\delta (\Delta Q)}{\Delta Q_{_0}}\right| \leq 10\% 
\end{equation}
with $\Delta Q_{_0} \simeq 1.294$ MeV \cite{pdg}, or equivalently
\begin{equation}
\label{5}\delta (\Delta Q)\leq 0.129 \; {\rm MeV} 
\end{equation}

\vspace{0.7cm}

At this stage, we must work out the cosmological
solution for the radiation dominated era. We shall consider 
the warped product $M^4 \times R$,
where $M^4$ is the ordinary 4-dimensional spacetime manifold and $R$
correspond to the extra dimension. This leads to the line element
\begin{equation}
ds^2 = - dt^2 \,+ \,\frac{A^2 (t)}{(1+k(x_{_1}^2 + x_{_2}^2 +
x_{_3}^2)/4)^2} (\, dx_{_1}^2 \,+\, dx_{_2}^2 \,+\, dx_{_3}^2 \,) \,+\, e^{\zeta(t)} dx_{_5}^2
\end{equation}
where $A^2(t)$ is the expansion scale factor and $e^{\zeta (t)}$ must
be associated with the mass scale factor. This could be simplified
considering a flat spatial section in $M^4$ (i.e.
$k=0$). This assumption is justified when one compares the order of
magnitude of the different terms in the Einstein equation evaluated 
in the radiation era \cite{Blaq}.

The suitable generalization of Einstein equation for the theory can be
written as\footnote{In what follows, latin indices 
$A, B,\dots$, run from 0 to 4, greek indices from 0 to 3 and latin 
indices $i,
j, \dots$, from 1 to 3.}

\begin{mathletters}
\begin{equation}
G_{AB} = 8 \; \pi \; G \; \, T_{AB}
\label{E}
\end{equation}
\begin{equation}
T_{AB} = {\rm diag} ( \rho,\, -p,\, -p,\, -p,\, 0 )
\label{T}
\end{equation}
\end{mathletters}
with $\rho$ ($p$) the density (pressure) of the radiation fluid, and $G$ is 
Newton's gravitational constant. The equation of state is $p =
1/3 \rho$. The boundary condition that ought to be imposed is a smoothly
matching at $t = t_{\rm eq}$ ( equivalence time) with the
dust-filled solution obtained in a previous paper \cite{Luis}. 
It is important to stress that in our previous work
we required
that both $\zeta$ and $ d \zeta / dt $ vanish at $t$ = today.
With these conditions, the masses of the fundamental particles can be 
set to 
their present
experimental value \cite{pdg}, and mass variations are negligible 
in short 
time scales (see eq.(\ref{masa})), which is consistent with the bound
$\vert\dot{m}/m\vert _{\rm today}\alt 10^{-12}$ yr$^{-1}$ \cite{Pipi}. 
However, our results are to a large extent independent of the 
epoch in which the initial conditions for $\zeta$ and $d\zeta /dt$ 
were imposed. \footnote{This can be seen by comparing the prediction 
of this theory for
$\dot{m}/m=\dot{\zeta}/2$ at $t=$ today in the case where the 
initial
conditions for $\zeta$ and its derivative were imposed for instance at
the Earth formation epoch, with the current experimental limits
on $\dot{m}/m$ \cite{Pipi}.}

The radiation dominated era solution is given by
\begin{mathletters}
\begin{equation}
A^2 (t) =  2 \; \beta \; t 
\end{equation}
\begin{equation}
\zeta (t) = 2 \, \ln \left[  - \frac{1}{2} \left[ \frac{t_{\rm eq}}
{\sqrt{t_{_0}}}
- \sqrt{t_{_0}} \right] t^{-1/2} + \sqrt{\frac{t_{\rm eq}}{t_{_0}}} \,\right]
\label{op}
\end{equation}
\begin{equation}
\rho(t) = \frac{1}{16\,\pi\, G} \left\{ \frac{3}{2\,t^2} + \frac{3}{2\,t} 
\; \frac{\frac{1}{4}\; [\frac{t_{\rm eq}}{ \sqrt{t_{0}}} - \sqrt{t_0}\,] \; 
t^{-3/2}}{- \frac{1}{2} \;  
[\frac{t_{\rm eq}}{ \sqrt{t_{0}}} - \sqrt{t_0}\,] \; t^{-1/2} + 
\sqrt{\frac{t_{\rm eq}}{ t_0}}} \right\}
\end{equation}
(with $\beta$ a constant)\footnote{This solution was
previously obtained by Mann and Vincent \cite{Mann} imposing different 
boundary 
conditions. It was also obtained by Gr{\o}n \cite{Gron}; from his 
work it is clear that the rate of change of the fifth 
metric coefficient depends on initial conditions.}. 
\end{mathletters}

The way in which the mass is introduced in this formalism has an inherent 
ambiguity. The proposal of Ma \cite{Ma}
\begin{equation}
m(t) = \frac{c^2}{G} \int^{x_{_5} +\; l}_{_{x_5}} \, \sqrt{g_{55}} \; dx^5
\end{equation}
or, in the 
case of an
$x_5$-independent metric, 
\begin{equation}
m(\tau) = \frac{c^2}{G}\sqrt{g_{55}} \,\,\Delta l_0
\end{equation}
($\Delta l_0$ is the -finite- ``length'' of the body in the $x_5$
direction\cite{Ma}), does not specify the tensorial character of the
mass, which is implicitly introduced into the theory. In other
words, there is no conclusive reason to mantain the covariant form of
$g_{55}$ under the square root, and so $m(t)$ can be also scale as
\begin{equation}
m(\tau) = \frac{c^2}{G}\sqrt{g^{55}} \,\,\Delta \tilde{l}_0
\label{masa}
\end{equation}   
The bottom line in this idea is that the dimensional analysis used to
define the relationship between the mass and the extra dimension
\mbox{$x^5 = c^2 \, m\, / \, G$} does not exclude the covariant
formulation \mbox{$x_5 = c^2 \, m\, / \, G$}. We shall adopt
hereafter the most favorable definition for the theory.

Hence, we can estimate the variation of the mass from eq. (\ref{op}).
Since $t_0 \gg t_{\rm eq} \gg t_{\rm nuc}$ ($t_0$ is the present age
of the universe), we find
\begin{equation}
\zeta_{\rm nuc} \approx 2 \ln \frac{1}{2} \sqrt{\frac{t_0}{t_{\rm
nuc}}} \approx 2 \ln 10^8
\end{equation}
Defining the quotient $\Delta m / m$ as
\begin{equation}
\frac{\Delta m}{m_0}=\frac{ m (t_{_{\rm nuc}}) - m (t_{_0}) } 
{ m(t_{_0}) } = e^{-\zeta_{_{\rm nuc}}/2} - 1 \, \simeq \, -1
\label{coc}
\end{equation}
we are able to see that the theory predicts 
\begin{equation}
\left|\frac{\delta \Delta Q}{\Delta Q_{_0}}\right| \simeq 100 \%
\end{equation}
which is in disagreement with the previous bound (\ref{4}). Using a covariant 
formulation for the mass scale factor in (\ref{coc}) one would get an 
even worse disagreement.

Thereupon, the assumed relation connecting the 5$^{\rm th}$ dimension
and the particles rest masses is false, at least in the case we have 
explored. More general (i.e. $x_5$ dependant) metrics deserve more
thorough analysis. We hope to report on this issues in a forthcoming work. 

\vspace{1cm}

We wish to thank Santiago Perez Bergliaffa and Harald Soleng for 
the critical reading of the manuscript and valuable comments. Remarks
by an anonymous referee are gratefully acknowledged. This work  
has been partially supported by CONICET, UNLP and Fundaci\'on Antorchas.

\end{document}